\documentclass[aps,preprint]{revtex4-1}%
\usepackage{bm}
\usepackage{amsfonts}
\usepackage{amsmath}
\usepackage{amssymb}
\usepackage{graphicx}%
\begin{document}

\title{Zero sound in triplet-correlated superfluid neutron matter}
\author{L. B. Leinson}

\begin{abstract}
The linear response of a superfluid neutron liquid onto external vector field
is studied for the case of $^{3}P_{2}-\,^{3}F_{2}$ pairing. In particular, we
analyze the case of neutron condensation into the state with $m_{j}=0$ which
is conventionally considered as the preferable one in the bulk matter of
neutron stars. Consideration is limited to the case when the wave-length of a
perturbation is large as compared to the coherence length in the superfluid
matter and the transferred energy is small in comparison with the gap
amplitude. The obtained results are used to analyse collisionless sound-like
excitations of the superfluid condensate. Zero sound (if it exists) is found
to be anisotropic and undergoes strong decrement below some temperature
threshold depending substantially on the intensity of Fermi-liquid interactions.

\end{abstract}
\maketitle

\affiliation{Institute of Terrestrial Magnetism, Ionosphere and Radio Wave Propagation RAS,
142190 Troitsk, Moscow Region, Russia}

\startpage{1}

\section{Introduction}

At temperatures below the critical value $T_{c}$ the neutron Fermi liquid in
the bulk matter of neutron stars is expected to develop a triplet superfluid
condensate \cite{Tamagaki, Hoffberg}. Considerable work has done with the most
realistic nuclear potentials for determining the magnitude of the superfluid
gap at different matter densities \cite{Takatsuka, Baldo, Elg, Khod, Dean,
Khodel}, while the low-energy collective excitations of such superfluid liquid
are not well investigated as yet. In the meantime such excitations can play an
important role in the evolution of neutron stars. For example, the decay of
collective spin waves into neutrino pairs, occuring in a superfluid core of
neutron star through neutral weak currents, presents a new mechanism of
intensive cooling in some domain of low temperatures \cite{L10, L10a, L10b}. A
recent investigation \cite{Pons} has shown sound-like oscillations of
$^{1}S_{0}$ superfluid neutron matter (called superfluid phonons) due to a
very large mean-free-path influence heat conduction in a magnetized crust of
neutron stars, where the motion of electrons is very anisotropic. Analogous
effects might be expected in the neutron star core, since spontaneous breaking
of rotation invariance and the baryon number caused by the triplet
condensation should lead to the appearance of several Goldstone modes
\cite{Bed}.

Generally speaking under $\beta$-equilibrium the superdense core of neutron
stars is composed of neutrons with a small admixture of protons and electrons.
A fraction of hyperons can also appear at higher densities. It is well known
however that long-range electromagnetic interactions push out collective
oscillations of the charged particles up to the plasma frequency which is
large sufficiently for (approximate) decoupling of the plasma modes from the
sound-like oscillations of neutral component \cite{Nambu, LP06, L01}. In this
case the effect of short-range proton-neutron interactions is reduced mostly
to renormalization of the effective mass of neutrons participating in the
collective oscillations \cite{Migdal, Gusakov}. Therefore the problem can be
simplified considering the sound-like excitations in a pure neutron superfluid liquid.

Previously the sound modes at finite temperatures have been investigated for
isotropic singlet-spin superfluids \cite{Leggett} and for the case of triplet
$p$-wave pairing in superfluid liquid $^{3}He$ \cite{Wolfe73, Wolfe, Maki}.
Although a qualitative picture of the sound-like waves in superfluids is very
similar, the above theories cannot be immediately applied to the case of
superdense superfluid neutron matter. The well-developed theory of isotropic
pairing cannot be applied because the triplet condensate in superdense nuclear
matter is expected to be anisotropic. The theory of sound-like collective
excitations in an anisotropic phase of superfluid $^{3}He$ is designed only
for extremely fast waves with a velocity that is very large as compared to the
Fermi velocity.

Since spin-orbit and tensor interactions between neutrons are known to
dominate at high densities the neutron pairing involves a mixing of $^{3}%
P_{2}$ and $^{3}F_{2}$ channels \cite{Takatsuka, Baldo, Elg, Khod, Dean,
Khodel}. The sum of spin-orbit and tensor interactions cannot be described
with the aid of a sole coupling constant. This complicates the investigation
of collective excitations in standard ways with making use of an explicit form
of the pairing interaction, where the sole coupling constant drops out of the
equations by virtue of the gap equation.

However, when the wave-length of the perturbation is large as compared to the
coherence length in the superfluid matter and the transferred energy is small
in comparison with the gap amplitude, as is typical for sound-like
excitations, the collective motion of the condensate can be described in terms
of total variable phase which (in the BCS approximation) can be derived
immediately from the current conservation condition. This approach, for the
first time suggested in Ref. \cite{L08}, allows to avoid any explicit form of
the interaction in the pairing channel. Residual Fermi-liquid interactions can
be incorporated into the theory as a set of molecular fields \cite{Leggett75}.
In application to polarization functions, this approach is well developed in
Ref. \cite{Gusakov}.

This paper is organized as follows. Section II contains some preliminary notes
on how the Fermi-liquid interactions can be reduced to molecular fields. In
Sec. III we derive, in the BCS approximation, the linear response of the
triplet-correlated superfluid neutron liquid onto an effective vector field
given by the sum of external and molecular fields. In Sect. IV we express
self-consistently the effective fields via external fields, thus obtaining the
linear medium response by taking into account the Fermi-liquid interactions.
In Sect. V we analyze the poles of the longitudinal response function in order
to derive the dispersion of sound-like oscillations in the condensate. Section
VI contains a short summary of our findings and the conclusion. Throughout
this paper, we use the system of units $\hbar=c=1$, and the Boltzmann constant
$k_{B}=1$.

\section{Fermi-liquid interactions and molecular fields}

It is well known that the Landau theory of a normal Fermi-liquid is based on
the fact that a large part of the interactions can be taken into account with
the aid of renormalizations effects. An effective Hamiltonian of the system
contains the renormalized single-particle energy of quasiparticles with
occupation numbers $n(\mathbf{{p},\sigma)}$ and a residual interaction between
changes $\delta n$ in the quasiparticle occupation at the Fermi surface. As
has been shown by Leggett \cite{Leggett65}, the quasiparticle pairing does not
change the net occupation for a given direction on the Fermi surface, if
approximate particle-hole symmetry is maintained. Thus the Fermi-liquid
interactions remain unchanged upon pairing. In other words, the Fermi-liquid
interactions do not interfere with the pairing phenomenon.

In our analysis we shall also assume that the anisotropy of the order
parameter, which takes place in the case of triplet-spin pairing, plays no
significant role in the Fermi-liquid interactions. This assumption is clearly
justified because the characteristic length associated with the Fermi-liquid
interactions is of the order of the inverse Fermi momentum, $p_{F}^{-1}$, and
hence is much smaller than any other characteristic length entering the
problem. This allows us to disregard the spin-dependent part of Fermi-liquid
interactions in the vector channel we shall consider.

Since we are interested in values of the neutron momenta near the Fermi
surface, $\mathbf{p}\simeq p_{F}\mathbf{n}$, the amplitudes of the
Fermi-liquid interactions $f\left(  \mathbf{nn}^{\prime}\right)  $ can be
expanded into Legendre polynomials. In the vector channel these interactions
are spin-independent and can be completely described in terms of the infinite
set of Landau parameters $F_{l}$. In practice even for a saturated nuclear
matter one does not know the Landau parameters $F_{l}$ for $l\geq2$, and in
actual calculations they are frequently put equal zero. The remaining
Fermi-liquid interactions can be written in the form
\begin{equation}
\varrho f\left(  \mathbf{nn}^{\prime}\right)  =F_{0}+F_{1}\mathbf{nn}^{\prime
},\label{ph2}%
\end{equation}
where $\varrho=p_{F}M^{\ast}/\pi^{2}$ is the density of states near the Fermi
surface in the normal state, and the effective mass of a neutron quasiparticle
is defined as $M^{\ast}=p_{F}/\upsilon_{F}$, where $\upsilon_{F}$ is the Fermi velocity.

This approach can be considered as a model of the Fermi-liquid interactions.
It is known, however, that the Landau interactions with $l\geq2$ do not modify
the longitudinal response functions of normal (nonsuperfluid) Fermi liquid.
For a one-component Fermi liquid this was demonstrated, for example, in Ref.
\cite{Pines}. The same result was obtained for a one-component $^{1}S_{0}$
superfluid Fermi liquid in Ref. \cite{Larkin} where the effective vertices and
the polarization functions (at $\omega,q\upsilon_{F}\ll\Delta$) have been
found to depend only on $F_{0}$ and $F_{1}$.

Interactions (\ref{ph2}) renormalize the normal energy of a quasiparticle in
the weak external vector field $A^{\mu}=\left(  A_{0},\mathbf{A}\right)  $ as%
\begin{equation}
\tilde{\varepsilon}\left(  \mathbf{p}\right)  \simeq\varepsilon\left(
\mathbf{p}\right)  +c_{V}A_{0}-\frac{c_{V}}{M}\mathbf{pA}+\frac{1}{\varrho
}F_{0}\sum_{p,\sigma}\delta n\left(  \mathbf{p}^{\prime},\sigma^{\prime
}\right)  +\frac{1}{\varrho p_{F}^{2}}F_{1}\mathbf{p}\sum_{p^{\prime},\sigma
}\mathbf{p}^{\prime}\delta n\left(  \mathbf{p}^{\prime},\sigma\right)
~.\label{ep}%
\end{equation}
We denote as $\varepsilon_{\mathbf{p}}=\upsilon_{F}\left(  p-p_{F}\right)  $
the quasiparticle energy related to the Fermi energy in the normal state;
$c_{V}$ is the coupling constant, which depends on the nature of the external
field, and $n\left(  \mathbf{p}^{\prime},\sigma^{\prime}\right)  $ is the
distribution function of neutron quasiparticles with momenta $\mathbf{p}$ and
spin $\sigma$.

From Eq. (\ref{ep}) it is clearly seen that Fermi-liquid interactions can be
reduced to molecular fields \cite{Leggett75}, defined as
\begin{equation}
A_{0}^{mol}\equiv\frac{1}{c_{V}\varrho}F_{0}\sum_{p,\sigma}\delta n\left(
\mathbf{p},\sigma\right)  ~, \label{A0mol}%
\end{equation}
and%
\begin{equation}
\mathbf{A}^{mol}\equiv-\frac{F_{1}M}{c_{V}\varrho p_{F}^{2}}\mathbf{\sum
_{p,\sigma}p}\delta n\left(  \mathbf{p},\sigma\right)  ~. \label{Amol}%
\end{equation}
Then Eq. (\ref{ep}) can be reduced to the form%
\begin{equation}
\tilde{\varepsilon}\left(  \mathbf{p}\right)  =\varepsilon\left(
\mathbf{p}\right)  +c_{V}\left(  A_{0}^{eff}-\frac{\mathbf{p}}{M^{\ast}%
}\mathbf{A}^{eff}\right)  ~, \label{etilde}%
\end{equation}
where the effective fields are given by the sum of external and molecular
fields,
\begin{equation}
A_{0}^{eff}=A_{0}+A_{0}^{mol},\mathbf{A}^{eff}=\frac{M^{\ast}}{M}\left(
\mathbf{A}+\mathbf{A}^{mol}\right)  ~. \label{aef}%
\end{equation}

The molecular fields depend on the charge perturbation and current density and
should be calculated consistently. Therefore first we shall perform the
calculation of the medium response onto the effective field and next find the
explicit form of the effective fields in a self-consistent way.

\section{BCS response in the limit of $\omega,q\upsilon_{F}\ll\Delta$}

The triplet order parameter in the neutron superfluid is a symmetric matrix in
spin space which can be written as
\begin{equation}
\mathcal{\hat{D}}\left(  \mathbf{n}\right)  =\Delta\mathbf{\bar{b}}\left(
\mathbf{n}\right)  \bm{\hat{\sigma}}\hat{g}~,\label{Dn}%
\end{equation}
where $\hat{\bm{\sigma}}=\left(  \hat{\sigma}_{1},\hat{\sigma}_{2},\hat
{\sigma}_{3}\right)  $ are Pauli spin matrices; $\hat{g}\equiv i\hat{\sigma
}_{2}$, with $\hat{g}\hat{g}=-\hat{1}$; and $\hat{1}$ is the $2\times2$ unit
matrix in spin space. In the ground state, the gap amplitude $\Delta$ is a
constant (on the Fermi surface), and $\mathbf{\bar{b}}\left(  \mathbf{n}%
\right)  $ is a real vector in spin space which we normalize by the condition
$\left\langle \bar{b}^{2}\left(  \mathbf{n}\right)  \right\rangle =1$.
Hereafter the angle brackets denote angle averages, $\left\langle
...\right\rangle \equiv\left(  4\pi\right)  ^{-1}\int d\mathbf{n}...~$. The
angular dependence of the order parameter is represented by the unit vector
$\mathbf{n=p}/p$ which defines the polar angles $\left(  \theta,\varphi
\right)  $ on the Fermi surface. In the components, $n_{1}=\sin\theta
\cos\varphi,~n_{2}=\sin\theta\sin\varphi,~n_{3}=\cos\theta$.

Making use of the adopted graphical notation for the ordinary and anomalous
propagators, $\hat{G}=\parbox{1cm}{\includegraphics[width=1cm]{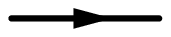}}$,
$\hat{G}^{-}(p)=\parbox{1cm}{\includegraphics[width=1cm,angle=180]{Gn.eps}}$,
$\hat{F}^{(1)}=\parbox{1cm}{\includegraphics[width=1cm]{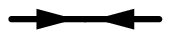}}$\thinspace,
and $\hat{F}^{(2)}=\parbox{1cm}{\includegraphics[width=1cm]{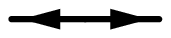}}$%
\thinspace, it is convenient to employ the Matsubara calculation technique for
the system in thermal equilibrium. Then the analytic form of the propagators
is as follows (see, \textit{e.g.}, Ref. \cite{L10})
\begin{align}
\hat{G}\left(  \eta_{n},\mathbf{p}\right)   &  =G\left(  \eta_{n}%
,\mathbf{p}\right)  \delta_{\alpha\beta}~,\ \ \ \ \ \ \ \hat{G}^{-}\left(
\eta_{n},\mathbf{p}\right)  =G^{-}\left(  \eta_{n},\mathbf{p}\right)
\delta_{\alpha\beta}~,\nonumber\\
\hat{F}^{\left(  1\right)  }\left(  \eta_{n},\mathbf{p}\right)   &  =F\left(
\eta_{n},\mathbf{p}\right)  \mathbf{\bar{b}}\bm{\hat{\sigma}}\hat
{g}~,\ \ \ \hat{F}^{\left(  2\right)  }\left(  \eta_{n},\mathbf{p}\right)
=F\left(  \eta_{n},\mathbf{p}\right)  \hat{g}\bm{\hat{\sigma}}\mathbf{\bar{b}%
}~,\label{GF}%
\end{align}
where the scalar Green's functions are of the form $G^{-}\left(  \eta
_{n},\mathbf{p}\right)  =G\left(  -\eta_{n},-\mathbf{p}\right)  $ and%
\begin{equation}
G\left(  \eta_{n},\mathbf{p}\right)  =\frac{-i\eta_{n}-\varepsilon
_{\mathbf{p}}}{\eta_{n}^{2}+E_{\mathbf{p}}^{2}}~,\ F\left(  \eta
_{n},\mathbf{p}\right)  =\frac{\Delta}{\eta_{n}^{2}+E_{\mathbf{p}}^{2}%
}~.\label{GFc}%
\end{equation}
In the above, $\eta_{n}\equiv\pi\left(  2n+1\right)  T$ with $n=0,\pm
1,\pm2,...$ is Matsubara's fermion frequency, and we assume the "unitary gap
matrix" in the ground state, $\mathcal{\hat{D}}\left(  \mathbf{n}\right)
\mathcal{\hat{D}}^{\dagger}\left(  \mathbf{n}\right)  \propto\hat{1}$, thus
obtaining the energy of a one-particle excitation in the form $E_{\mathbf{p}%
}^{2}=\varepsilon_{\mathbf{p}}^{2}+\Delta^{2}\bar{b}^{2}\left(  \mathbf{n}%
\right)  $, where the (temperature-dependent) energy gap, $\Delta_{\mathbf{n}%
}=\Delta\left(  T\right)  \bar{b}\left(  \mathbf{n}\right)  $, is anisotropic.

The following notation will be used below. We designate as $L_{X,X}\left(
\omega,\mathbf{q;p}\right)  $ the analytical continuation of the Matsubara
sums:
\begin{equation}
L_{XX^{\prime}}\left(  \omega_{m},\mathbf{p+}\frac{\mathbf{q}}{2}%
\mathbf{;p-}\frac{\mathbf{q}}{2}\right)  =T\sum_{n}X\left(  \eta_{n}%
+\omega_{m},\mathbf{p+}\frac{\mathbf{q}}{2}\right)  X^{\prime}\left(  \eta
_{n},\mathbf{p-}\frac{\mathbf{q}}{2}\right)  ~, \label{LXX}%
\end{equation}
where $X,X^{\prime}\in G,F,G^{-}$, and $\omega_{m}=2\pi Tm$ with $m=0,\pm
1,\pm2...$.

It is convenient to divide the integration over the momentum space into an
integration over the solid angle $d\mathbf{n}$ and over the energy
$d\varepsilon_{\mathbf{p}}$ and operate with integrals%
\begin{equation}
\mathcal{I}_{XX^{\prime}}\left(  \omega,\mathbf{n,q};T\right)  \equiv\frac
{1}{2}\int_{-\infty}^{\infty}d\varepsilon_{\mathbf{p}}L_{XX^{\prime}}\left(
\omega,\mathbf{p+}\frac{\mathbf{q}}{2}\mathbf{,p-}\frac{\mathbf{q}}{2}\right)
~.\label{IXX}%
\end{equation}
These are functions of $\omega$, $\mathbf{q}$ and the direction of the
quasiparticle momentum $\mathbf{p}=p\mathbf{n}$. In deriving Eq. (\ref{IXX})
integration over $d\varepsilon_{\mathbf{p}}$ is extended to $-\infty$ since
the neutron matter is extremely degenerate.

Consider the medium response onto the effective vector field (\ref{aef}) in
the BCS approximation. In this case the ordinary three-point vector vertices
of a quasiparticle and a hole are defined in accordance with Eq.
(\ref{etilde}):
\begin{equation}
\gamma^{\mu}\left(  \mathbf{p}\right)  =\left(  1,\mathbf{p/}M^{\ast}\right)
~,~\gamma_{\mu}^{-}\left(  \mathbf{p}\right)  =\gamma_{\mu}\left(
-\mathbf{p}\right)  ~.\label{gmu}%
\end{equation}
We use greek letters for Dirac indices, $\mu=0,1,2,3$.

Variation of the anomalous self-energy $\mathcal{\hat{D}}\left(
\mathbf{n}\right)  $ in the field $A_{eff}^{\mu}$, can be described with the
aid of anomalous three-point vertices $\hat{T}_{\mu}^{\left(  1,2\right)  }$,
defined as:
\begin{equation}
c_{V}\hat{T}_{\mu}^{\left(  1\right)  }=\delta\mathcal{\hat{D}}^{\left(
1\right)  }/\delta A_{eff}^{\mu}~,~c_{V}\hat{T}_{\mu}^{\left(  2\right)
}=\delta\mathcal{\hat{D}}^{\left(  2\right)  }/\delta A_{eff}^{\mu
}~.\label{T12}%
\end{equation}
The anomalous vertices are $2\times2$ matrices in spin space which, near the
Fermi surface, depend on the transferred energy momentum $k^{\mu}=\left(
\omega,\mathbf{q}\right)  $ and the direction $\mathbf{n}=\mathbf{p}/p$ of the
quasiparticle velocity.

The Ward identity implies the following relations between the anomalous
vertices and the order parameter in the system \cite{Nambu, Schr} (see also
Refs. \cite{Migdal, L08}):
\begin{equation}
k^{\mu}T_{\mu}^{\left(  1\right)  }=2\mathcal{\hat{D}}\left(  \mathbf{n}%
\right)  ~,~k^{\mu}T_{\mu}^{\left(  2\right)  }=-2\mathcal{\hat{D}}^{\dagger
}\left(  \mathbf{n}\right)  ~.\label{Ward}%
\end{equation}

We now restrict our consideration to the case, when the wave-length of the
perturbation is large as compared to the coherence length and the transferred
energy is small in comparison with the gap amplitude, $\omega,q\upsilon_{F}%
\ll\Delta$. The only possible collective motion of the condensate in this case
is a variation of the total phase without a change of the order parameter
structure. Then the Ward identity reveals that for a uniform medium the
anomalous vertices can be written in the form
\begin{equation}
\hat{T}_{\mu}^{\left(  1\right)  }=Q_{\mu}\left(  \omega,\mathbf{q}\right)
\mathbf{\bar{b}}\bm{\hat{\sigma}}\hat{g}~,~\hat{T}_{\mu}^{\left(  2\right)
}=-Q_{\mu}\left(  \omega,\mathbf{q}\right)  \hat{g}%
\bm{\hat{\sigma}}\mathbf{\bar{b}}~,\label{T1}%
\end{equation}
where the unknown vector function $Q_{\mu}\left(  \omega,\mathbf{q}\right)  $
satisfies the condition%
\begin{equation}
k^{\mu}Q_{\mu}=2\Delta~.\label{QW}%
\end{equation}

For further progress, let us consider the retarded BCS polarization tensor in
the vector channel $\Pi^{\mu\nu}(\omega,\mathbf{q})$. The latter can be found
using the fact that the current in the system, $j^{\mu}\equiv\left(
\delta\rho,\mathbf{j}\right)  $ is connected to the linear correction
$\delta\hat{G}$ to the Green's function of a quasiparticle in the effective
external field $A_{eff}^{\mu}$ and can be obtained by analytic continuation of
the following Matsubara sums
\begin{equation}
\delta\rho=c_{V}T\sum_{p_{m}}\int\frac{d^{3}p}{8\pi^{3}}\mathrm{Tr}\left(
\delta\hat{G}\right)  ~,\label{ro}%
\end{equation}%
\begin{equation}
\mathbf{j}=\frac{c_{V}}{M^{\ast}}T\sum_{p_{m}}\int\frac{d^{3}p}{8\pi^{3}%
}\mathbf{p}\mathrm{Tr}\left(  \delta\hat{G}\right)  -\frac{Nc_{V}^{2}}%
{M^{\ast}}\mathbf{A}^{eff}\mathbf{~,}\label{j}%
\end{equation}
where $N=p_{F}^{3}/\left(  3\pi^{2}\right)  $\ is the total number density of neutrons.

\begin{figure}[h]
\includegraphics{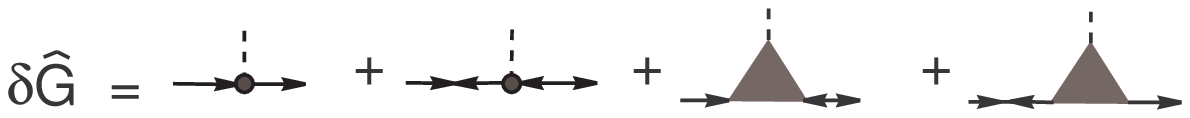}\caption{Correction to the ordinary propagator
of a neutron quasiparticle in the external field. The ordinary and anomalous
tree-point vertices are depicted by small circles and shadowed triangles,
respectively.}%
\label{fig1}%
\end{figure}

The linear correction to the Green's function of a quasiparticle caused by the
external field $A_{eff}^{\mu}$ is given by the diagrams shown in Fig.
\ref{fig1}, and can be written analytically as
\begin{align}
\delta\hat{G} &  =c_{V}\left(  GG\hat{\gamma}_{\mu}-FF~\hat{\gamma}_{\mu}%
^{-}\left(  \mathbf{\sigma\bar{b}}\right)  \left(  \mathbf{\sigma\bar{b}%
}\right)  \right.  \nonumber\\
&  \left.  +GF~\hat{T}_{\mu}^{\left(  1\right)  }\hat{g}\left(  \mathbf{\sigma
}\bar{\mathbf{b}}\right)  +FG~\left(  \mathbf{\sigma\bar{b}}\right)  \hat
{g}\hat{T}_{\mu}^{\left(  2\right)  }\right)  A_{eff}^{\mu}~,\label{dG}%
\end{align}
where the anomalous vertices are to be taken in the form (\ref{T1}), and we
use the notation $GG\equiv G\left(  \eta_{n}+\omega_{m},\mathbf{p+q}/2\right)
G\left(  \eta_{n},\mathbf{p-q}/2\right)  $, etc.,

Inserting Eq. (\ref{dG}) into Eqs. (\ref{ro}) and (\ref{j}) we can derive the
retarded polarization tensor $\Pi^{\mu\nu}$ with the aid of the standard
relation $j^{\mu}=\Pi^{\mu\nu}(\omega,\mathbf{q})A_{\nu}^{eff}$. In this way
after a little algebra we find%
\begin{equation}
\Pi_{\lambda\mu}=\Lambda_{\lambda\mu}+\Lambda_{\lambda}Q_{\mu}~,\label{Pi}%
\end{equation}
where $Q_{\mu}\left(  \omega,\mathbf{q}\right)  $ is still an unknown vector
to be found. The functions $\Lambda_{\lambda\mu}\left(  \omega,\mathbf{q}%
\right)  $ and $\Lambda_{\lambda}\left(  \omega,\mathbf{q}\right)  $ are given
by
\begin{equation}
\Lambda_{\lambda\mu}=c_{V}^{2}\varrho\left(  \left\langle \gamma_{\lambda
}\gamma_{\mu}\mathcal{I}_{GG}-\gamma_{\lambda}\gamma_{\mu}^{-}\bar{b}%
^{2}\mathcal{I}_{FF}\right\rangle +\frac{\upsilon_{F}^{2}}{3}\delta_{\lambda
i}\delta_{i\mu}\right)  ~,\label{Lmn}%
\end{equation}
and%
\begin{equation}
\Lambda_{\lambda}=c_{V}^{2}\varrho\left\langle 2\gamma_{\lambda}\bar{b}%
^{2}\mathcal{I}_{FG}\right\rangle =c_{V}^{2}\varrho\frac{1}{\Delta
}\left\langle \gamma_{\lambda}\left(  \omega+\mathbf{qv}\right)  \bar{b}%
^{2}\mathcal{I}_{FF}\right\rangle ~.\label{Lm}%
\end{equation}
To obtain the second equality we made use of the identity
\begin{equation}
\mathcal{I}_{FG}=-\mathcal{I}_{GF}=\frac{\omega+\mathbf{qv}}{2\Delta
}\mathcal{I}_{FF}~,\label{ident}%
\end{equation}
which can be verified by a straightforward calculation.

The function $Q_{\mu}\left(  \omega,\mathbf{q}\right)  $ can be found from the
requirement that the polarization tensor (\ref{Pi}) must satisfy current
conservation conditions, $\Pi^{\mu\nu}k_{\nu}=0$, and $k_{\nu}\Pi^{\nu\mu}=0$,
which can be written as the two coupled equations%
\begin{equation}
\Lambda^{\mu\nu}k_{\nu}+2\Delta\ \Lambda^{\mu}=0~,\label{Lmnq}%
\end{equation}%
\begin{equation}
k_{\nu}\Lambda^{\nu\mu}+k_{\nu}\Lambda^{\nu}\left(  q\right)  Q^{\mu}\left(
q\right)  =0~.\label{qLmn}%
\end{equation}
We made use of relation (\ref{QW}) in the first equation.

From the explicit form (\ref{Lmn}) for $\Lambda^{\mu\nu}$ one can easily find
the following relation
\begin{equation}
q^{\mu}\Lambda_{\mu\lambda}=\Lambda_{\lambda\mu}q^{\mu}+c_{V}^{2}%
\varrho\left\langle \left(  \gamma_{\lambda}\left(  \omega+\mathbf{qv}\right)
-\left(  \omega-\mathbf{qv}\right)  \hat{\gamma}_{\lambda}^{-}\right)  \bar
{b}^{2}\mathcal{I}_{FF}\right\rangle ~.\label{qmLmn}%
\end{equation}
The first term on the right-hand side of this expression can be replaced as
$\Lambda^{\mu\nu}k_{\nu}\rightarrow-2\Delta\ \Lambda^{\mu}$, in accordance
with Eq. (\ref{Lmnq}). Inserting the obtained result into Eq. (\ref{qLmn}), we
arrive at%
\begin{equation}
Q_{\lambda}=\frac{2\Delta\Lambda_{\lambda}-c_{V}^{2}\varrho\left\langle
\left(  \gamma_{\lambda}\left(  \omega+\mathbf{qv}\right)  -\left(
\omega-\mathbf{qv}\right)  \hat{\gamma}_{\lambda}^{-}\right)  \bar{b}%
^{2}\mathcal{I}_{FF}\right\rangle }{q^{\mu}\Lambda_{\mu}}~.\label{Qlam}%
\end{equation}
This formula can be simplified making use of the explicit form (\ref{Lm}) for
$\Lambda_{\lambda}$. In the components we obtain the expressions:
\begin{equation}
Q_{0}=2\Delta\frac{\omega\left\langle \bar{b}^{2}\mathcal{I}_{FF}\right\rangle
}{\left\langle \left(  \omega^{2}-\left(  \mathbf{qv}\right)  ^{2}\right)
\bar{b}^{2}\mathcal{I}_{FF}\right\rangle }~,\label{Q0}%
\end{equation}%
\begin{equation}
\mathbf{Q}=2\Delta\frac{\left\langle \mathbf{v}\left(  \mathbf{qv}\right)
\bar{b}^{2}\mathcal{I}_{FF}\right\rangle }{\left\langle \left(  \omega
^{2}-\left(  \mathbf{qv}\right)  ^{2}\right)  \bar{b}^{2}\mathcal{I}%
_{FF}\right\rangle }~,\label{Qv}%
\end{equation}
valid for $\omega,q\upsilon_{F}\ll\Delta$.

From Eqs. (\ref{Pi})--(\ref{Lm}), (\ref{Q0}) and (\ref{Qv}) one can obtain the
complete BCS polarization tensor $\Pi^{\mu\nu}\left(  \omega,\mathbf{q}%
\right)  $ in the vector channel. Due to conservation of the vector current
the polarization tensor can be decomposed into the sum of longitudinal (with
respect to $\mathbf{q}$) and transverse components, where the longitudinal and
transverse polarization functions are defined as $\Pi_{L}\equiv\Pi_{00}%
$,\ $\Pi_{T}\equiv\frac{1}{2}\left(  \delta_{ij}-q_{i}q_{j}/q^{2}\right)
\Pi_{ij}$, with $i,j=1,2,3$.

\section{Fermi-liquid effects}

We now turn to the Fermi-liquid effects. Our aim is to express the effective
fields (\ref{aef}) via external fields. For this we can use the fact that the
density current commutes with the bare interactions and does not need to be
renormalized. The current connection with the external field $\mathbf{A}$ is
given by the well-known relation
\begin{equation}
\sum_{p,\sigma}\mathbf{p}\delta n\left(  \mathbf{p},\sigma\right)  =\frac
{M}{c_{V}}\left(  \mathbf{j}+\frac{Nc_{V}^{2}}{M}\mathbf{A}\right)
\mathbf{~.}\label{Sum}%
\end{equation}
Inserting this into Eqs. (\ref{A0mol}) and (\ref{Amol}) we obtain the time
component of the effective field (\ref{aef}) in the form
\begin{equation}
A_{0}^{eff}\equiv A_{0}+\frac{1}{c_{V}^{2}\varrho}F_{0}\delta\rho
~,\label{A0ef}%
\end{equation}
Using also the continuity equation, $qj_{L}=\omega\rho$, we obtain the space
components of the effective field:
\begin{equation}
A_{L}^{eff}=A_{L}-F_{1}\frac{MM^{\ast}}{\varrho p_{F}^{2}c_{V}^{2}}%
\frac{\omega}{q}\delta\rho\label{ALef}%
\end{equation}%
\begin{equation}
\mathbf{A}_{T}^{eff}=\mathbf{A}_{T}-F_{1}\frac{MM^{\ast}}{\varrho p_{F}%
^{2}c_{V}^{2}}\mathbf{j}_{T}\label{ATef}%
\end{equation}

Insertion of the effective fields (\ref{A0ef}) and (\ref{ATef}) into
$\delta\rho=\Pi_{L}\left(  A_{0}^{eff}-\left(  \omega/q\right)  A_{L}%
^{eff}\right)  $ and $\mathbf{j}_{T}=-\Pi_{T}\mathbf{A}_{T}^{eff}$ after a
little algebra results in the following:
\begin{equation}
\delta\rho=\frac{\Pi_{L}}{1-\left(  F_{0}+s^{2}F_{1}/\left(  1+F_{1}/3\right)
\right)  c_{V}^{-2}\varrho^{-1}\Pi_{L}}\left(  A_{0}-\frac{\omega}{q}%
A_{L}\right)  ~,\label{dr}%
\end{equation}%
\begin{equation}
\mathbf{j}_{T}=-\frac{\Pi_{T}}{1-F_{1}/\left(  1+F_{1}/3\right)  \upsilon
_{F}^{-2}c_{V}^{-2}\varrho^{-1}\Pi_{T}}\mathbf{A}_{T}~.\label{jt}%
\end{equation}
In obtaining this we have used the Landau formula relating the bare mass of a
particle with a renormalized mass of a quasiparticle in the
translation-invariant system $M^{\ast}/M=1+F_{1}/3$.

The complete polarization functions $\tilde{\Pi}_{L,T}$ relate the density
perturbation and the density current with the external field as $\delta
\rho=\tilde{\Pi}_{L}\left( A_{0}- \left( \omega/q\right) A_{L}\right) $, and
$\mathbf{j}_{T}=-\tilde{\Pi}_{T}\mathbf{A}_{T}$, respectively. Then from Eqs.
(\ref{dr}) and (\ref{jt}) we obtain the complete polarization functions:%
\begin{equation}
\tilde{\Pi}_{L}=\frac{\Pi_{L}}{1-\left(  F_{0}+s^{2}F_{1}/\left(
1+F_{1}/3\right)  \right)  c_{V}^{-2}\varrho^{-1}\Pi_{L}}~, \label{PL}%
\end{equation}%
\begin{equation}
\tilde{\Pi}_{T}=\frac{\Pi_{T}}{1-F_{1}/\left(  1+F_{1}/3\right)  \upsilon
_{F}^{-2}c_{V}^{-2}\varrho^{-1}\Pi_{T}}~, \label{PT}%
\end{equation}
in agreement with the results of Gusakov \cite{Gusakov}.

Equations (\ref{PL}) and (\ref{PT}) completely describe the non-equilibrium
behavior of superfluid Fermi liquids in the vector-linear-response region for
not too high $\omega$ and $q$. Explicit expressions can be written using the
following notation:
\begin{equation}
\alpha\left(  s,\mathbf{h}\right)  =\frac{1}{2}+\frac{1}{2}\left\langle
\int_{0}^{\infty}d\varepsilon\frac{\left(  \cos^{2}\theta_{\mathbf{qn}}%
-s^{2}\right)  \varepsilon^{2}/E_{\mathbf{p}}^{2}}{s^{2}-\left(  \cos
^{2}\theta_{\mathbf{qn}}\right)  \varepsilon^{2}/E_{\mathbf{p}}^{2}}\frac
{dn}{dE_{\mathbf{p}}}\right\rangle ~, \label{alpha}%
\end{equation}%
\begin{equation}
\gamma\left(  s,\mathbf{h}\right)  =\frac{1}{2}\left\langle \int_{0}^{\infty
}d\varepsilon\frac{\cos\theta_{\mathbf{qn}}\left(  \cos^{2}\theta
_{\mathbf{qn}}-s^{2}\right)  \varepsilon^{2}/E_{\mathbf{p}}^{2}}{s^{2}-\left(
\cos^{2}\theta_{\mathbf{qn}}\right)  \varepsilon^{2}/E_{\mathbf{p}}^{2}}%
\frac{dn}{dE_{\mathbf{p}}}\right\rangle ~, \label{gamma}%
\end{equation}%
\begin{equation}
\zeta\left(  s,\mathbf{h}\right)  =\frac{1}{6}+\frac{1}{2}\left\langle
\int_{0}^{\infty}d\varepsilon\frac{\cos^{2}\theta_{\mathbf{qn}}\left(
\cos^{2}\theta_{\mathbf{qn}}-s^{2}\right)  \varepsilon^{2}/E_{\mathbf{p}}^{2}%
}{s^{2}-\left(  \cos^{2}\theta_{\mathbf{qn}}\right)  \varepsilon
^{2}/E_{\mathbf{p}}^{2}}\frac{dn}{dE_{\mathbf{p}}}\right\rangle ~.
\label{zeta}%
\end{equation}
\begin{equation}
\eta\left(  s,\mathbf{h}\right)  =\left\langle \int_{0}^{\infty}%
d\varepsilon\frac{\left(  \cos^{2}\theta_{\mathbf{qn}}\right)  \varepsilon
^{2}/E_{\mathbf{p}}^{2}}{s^{2}-\left(  \cos^{2}\theta_{\mathbf{qn}}\right)
\varepsilon^{2}/E_{\mathbf{p}}^{2}}\frac{dn}{dE_{\mathbf{p}}}\right\rangle ~,
\label{eta}%
\end{equation}%
\begin{equation}
\chi\left(  s,\mathbf{h}\right)  =\left\langle \int_{0}^{\infty}%
d\varepsilon\frac{s\left(  \cos\theta_{\mathbf{qn}}\right)  \varepsilon
^{2}/E_{\mathbf{p}}^{2}}{s^{2}-\left(  \cos^{2}\theta_{\mathbf{qn}}\right)
\varepsilon^{2}/E_{\mathbf{p}}^{2}}\frac{dn}{dE_{\mathbf{p}}}\right\rangle ~,
\label{khi}%
\end{equation}
where $s=\omega/\left(  q\upsilon_{F}\right)  $,
\begin{equation}
\frac{dn}{dE_{\mathbf{p}}}\equiv\frac{1}{2T}\cosh^{-2}\frac{E_{\mathbf{p}}%
}{2T}~, \label{dndE}%
\end{equation}
and we assume that the angle $\theta_{\mathbf{qn}}$ between the quasiparticle
momentum $\mathbf{p}=p\mathbf{n}$ and the momentum transfer $\mathbf{q}$ is
fixed by the relation%
\begin{equation}
\cos\theta_{\mathbf{qn}}=h_{x}\sin\theta\cos\varphi+h_{y}\sin\theta\sin
\varphi+h_{z}\cos\theta~, \label{nq}%
\end{equation}
where the unit vector $\mathbf{h\equiv q}/q=\left(  h_{x},h_{y},h_{z}\right)
$ defines the direction of the momentum transfer, and the unit vector
$\mathbf{n}=\left(  \sin\theta~\cos\varphi,\sin\theta~\sin\varphi,\cos
\theta\right)  $ defines the polar angles $\left(  \theta,\varphi\right)  $ on
the Fermi surface.

Then the longitudinal polarization function (\ref{PL}) can be written as
\begin{equation}
\tilde{\Pi}_{L}=c_{V}^{2}\varrho\frac{Q\left(  s,\mathbf{h}\right)
}{1-\left(  F_{0}+s^{2}F_{1}/\left(  1+F_{1}/3\right)  \right)  Q\left(
s,\mathbf{h}\right)  }~, \label{PLL}%
\end{equation}
where
\begin{equation}
Q\left(  s,\mathbf{h}\right)  =\eta\left(  s,\mathbf{h}\right)  +\chi\left(
s,\mathbf{h}\right)  +2\alpha\left(  s,\mathbf{h}\right)  \frac{s\gamma\left(
s,\mathbf{h}\right)  +\zeta\left(  s,\mathbf{h}\right)  }{s^{2}\alpha\left(
s,\mathbf{h}\right)  -\zeta\left(  s,\mathbf{h}\right)  }~. \label{PLBCSL}%
\end{equation}

\section{Sound-like excitations}

The pole of the density fluctuation propagator (\ref{PLL}) at%
\begin{equation}
\left(  F_{0}+s_{0}^{2}F_{1}/\left(  1+F_{1}/3\right)  \right)  Q\left(
s_{0},\mathbf{h};T\right)  =1~,\label{Qsh}%
\end{equation}
defines the dispersion, $s=s_{0}$, of the "collisionless" collective mode,
with $\omega,q\upsilon_{F}\ll\Delta$.

Equation (\ref{Qsh}) with $Q\left(  s,\mathbf{h}\right)  $ as given in Eq.
(\ref{PLBCSL}) generalizes previous results of Refs. \cite{Leggett} and
\cite{Wolfe73, Wolfe, Maki} to the case of pairing caused by  spin-orbit and
tensor interactions. Therefore before proceeding to the detailed analysis of
the sound propagation in the $^{3}P_{2}-\,^{3}F_{2}$ neutron superfluid, we
examine the obtained equations for the particular cases of the triplet-spin
condensate in superfluid $^{3}He$.

Consider first \textit{the case of isotropic pairing}. If the energy gap is
isotropic, the angle integrals in Eqs. (\ref{alpha})--(\ref{khi}) can be
performed by assuming the polar axis along the transferred momentum. We then
obtain $\gamma\left(  s,\mathbf{h}\right)  =\chi\left(  s,\mathbf{h}\right)
=0$ and the longitudinal polarization function reduces to Eq. (\ref{PLL})
with
\begin{equation}
Q\left(  s\right)  =\eta\left(  s\right)  +\frac{2\alpha\left(  s\right)
\zeta\left(  s\right)  }{s^{2}\alpha\left(  s\right)  -\zeta\left(  s\right)
}~,\label{QLs}%
\end{equation}
as obtained by Leggett \cite{Leggett} for the case of isotropic $s$-wave pairing.

Consider now \textit{the case of anisotropic pairing at} $T=0$. In this case
the quantities $\alpha\left(  s,\mathbf{h}\right)  \simeq1/2$ and
$\zeta\left(  s,\mathbf{h}\right)  =1/6$ are independent of $s$ and again
isotropic, while $\gamma\left(  s,\mathbf{h}\right)  =\eta\left(
s,\mathbf{h}\right)  =\chi\left(  s,\mathbf{h}\right)  =0$. Thus for $T=0$ we
find
\begin{equation}
\tilde{\Pi}_{L}=\frac{c_{V}^{2}\varrho}{\left(  1+F_{0}\right)  }\frac
{s_{0}^{2}}{s^{2}-s_{0}^{2}}~,~s_{0}^{2}=\frac{1}{3}\left(  1+F_{0}\right)
\left(  1+F_{1}/3\right)  ~,\label{T0vel}%
\end{equation}
in agreement with the result obtained by W\"{o}lfe \cite{Wolfe} for an
anisotropic phase of superfluid $^{3}He$ at zero temperature. Notice that the
same was obtained also by Leggett \cite{Leggett} for the case of isotropic
$s$-wave pairing. The pole at $s=s_{0}$ corresponds to the first sound
("Bogolyubov-Anderson" mode) undamped at zero temperature.

We turn now to \textit{the case of anisotropic} $p$-\textit{wave pairing in
liquid} $^{3}He$. The experimentally observable sound velocity in liquid
$^{3}He$ is large, therefore it is traditional to calculate the sound
dispersion in the limit $s\gg1$. Expanding Eq. (\ref{PLBCSL}) in powers of
$1/s$, one can obtain up to accuracy $s^{-4}$,%

\begin{equation}
Q\left(  s,\mathbf{h}\right)  \simeq\allowbreak\frac{5}{3}\frac{1}{5s^{2}%
-3}+\frac{1}{s^{4}}\left(  -\left\langle \lambda\left(  \mathbf{n}\right)
\cos^{4}\theta_{\mathbf{nq}}\right\rangle +\frac{\left\langle \lambda\left(
\mathbf{n}\right)  \cos^{2}\theta_{\mathbf{nq}}\right\rangle ^{2}}%
{\lambda\left(  \mathbf{n}\right)  }\right)  ~, \label{Qser}%
\end{equation}
where%
\begin{equation}
\lambda\left(  \mathbf{n}\right)  \equiv\int_{0}^{\infty}d\varepsilon
\frac{\Delta^{2}}{E^{3}}\tanh\frac{E}{2T}~. \label{lambda}%
\end{equation}
In obtaining Eq. (\ref{Qser}) we used the identity
\begin{equation}
\int_{0}^{\infty}d\varepsilon\frac{\varepsilon^{2}}{E_{\mathbf{p}}^{2}}%
\frac{dn}{dE_{\mathbf{p}}}=1-\int_{0}^{\infty}d\varepsilon\frac{\Delta^{2}%
}{E_{\mathbf{p}}^{3}}\tanh\frac{E_{\mathbf{p}}}{2T}~. \label{iden}%
\end{equation}
Inserting Eq. (\ref{Qser}) into Eq. (\ref{Qsh}) one can obtain the dispersion
law for high-frequency sound in superfluid $^{3}He$. Assuming $F_{0}\sim
s_{0}^{2}\gg1$ we find
\begin{align}
s_{0}^{2}\allowbreak &  =\frac{1}{3}\left(  F_{0}+\frac{9}{5}\right)  \left(
1+\frac{1}{3}F_{1}\right) \nonumber\\
&  +\left(  3+F_{1}\right)  \left(  -\left\langle \lambda\left(
\mathbf{n}\right)  \cos^{4}\theta_{\mathbf{nq}}\right\rangle +\frac
{\left\langle \lambda\left(  \mathbf{n}\right)  \cos^{2}\theta_{\mathbf{nq}%
}\right\rangle ^{2}}{\lambda\left(  \mathbf{n}\right)  }\right)  ~,
\label{HeA}%
\end{align}
in agreement with the expression derived by W\"{o}lfe \cite{Wolfe}.

We focus now on \textit{the sound propagation in the} $^{3}P_{2}-\,^{3}F_{2}$
\textit{superfluid neutron liquid} which is expected to exist in neutron stars
at supernuclear densities. First one has to specify the order parameter
(\ref{Dn}) for the particular case of neutron pairing. It is conventional to
represent the triplet order parameter of the system as a superposition of
standard spin-angle functions $\hat{\Phi}_{jlm_{j}}$ of the total angular
momentum $\left(  j=2,m_{j}\right)  $ with partial amplitudes $\Delta_{lm_{j}%
}$:
\begin{equation}
\hat{D}=\sum_{lm_{j}}\Delta_{lm_{j}}\hat{\Phi}_{jlm_{j}}\left(  \mathbf{n}%
\right)  ~.\label{1}%
\end{equation}
In our calculations we use vector notation which involves a set of mutually
orthogonal complex\ vectors $\mathbf{b}_{lm_{j}}\left(  \mathbf{n}\right)  $
defined as
\begin{equation}
\mathbf{b}_{lm_{j}}\left(  \mathbf{n}\right)  =-\left(  1/2\right)
\mathrm{Tr}\left(  \hat{g}\bm{\hat{\sigma}}\hat{\Phi}_{jlm_{j}}\right)
\label{2}%
\end{equation}
and normalized by the condition $\left\langle \mathbf{b}_{l^{\prime}%
m_{j}^{\prime}}^{\ast}\mathbf{b}_{lm_{j}}\right\rangle =\delta_{ll^{\prime}%
}\delta_{m_{j}m_{j}^{\prime}}$. We will focus on the $^{3}P_{2}-\,^{3}F_{2}$
condensation into the state with $m_{j}=0$ which is conventionally considered
as the preferable one in the bulk matter of neutron stars. In this case one
has%
\begin{equation}
\bar{b}^{2}\left(  \mathbf{n}\right)  =\frac{1}{2}\left(  1+3n_{3}^{2}\right)
\delta_{1}+\frac{3}{4}\left(  5n_{3}^{4}-2n_{3}^{2}+1\right)  \delta
_{3}~,\label{bsq}%
\end{equation}
where $\delta_{1}=\Delta_{1}^{2}/\Delta^{2}$ and $\delta_{3}=\Delta_{3}%
^{2}/\Delta^{2}$ are partial contributions of the ${^{3}P_{2}}$ and
${^{3}F_{2}}$ states, respectively, $\delta_{1}+\delta_{3}=1$.

For $0<T<T_{c}$ the behavior of $\tilde{\Pi}_{L}$ in the intermediate region
of $s$ depends essentially on the temperature. According to Eqs.
(\ref{alpha})-(\ref{khi}) the imaginary part of the functions arises from the
pole of the integrand at $s^{2}=\left(  \cos^{2}\theta_{\mathbf{qn}}\right)
\varepsilon^{2}/E_{\mathbf{p}}^{2}$. This is Cherenkov's condition which can
be satisfied only if $s<1$. Neglecting the narrow temperature domain where the
imaginary part of polarization is exponentially small \cite{Leggett}, one can
conclude that the well-defined (undamped) waves correspond to $s>1$. Further
we consider only undamped sound-like oscillations with $s>1$.
\begin{figure}[h]
\includegraphics{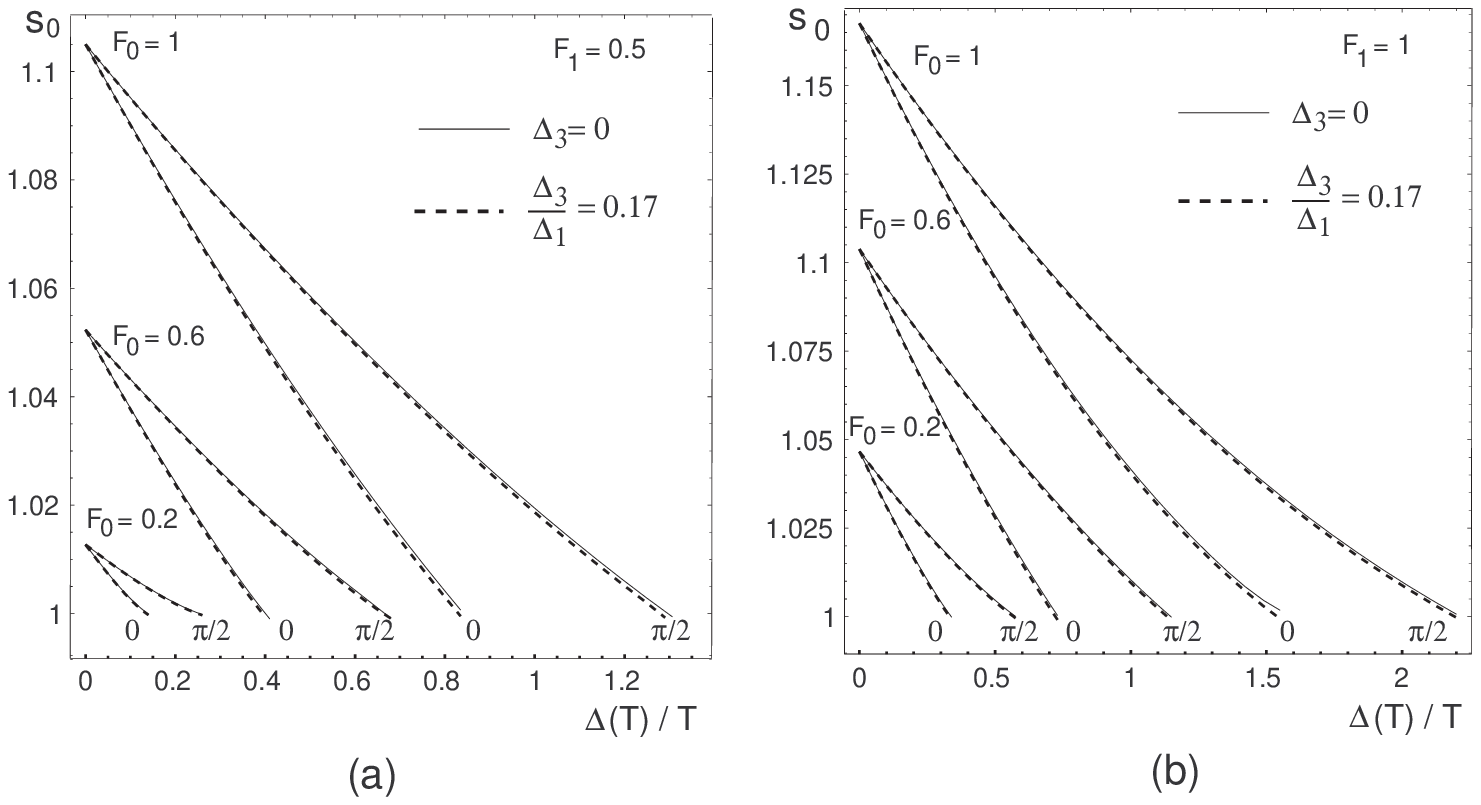}\caption{Zero-sound velocity as a function of
temperature below the critical point, $T<T_{c}$. The dimensionless velocity of
zero sound, $s_{0}=\omega_{0}/(q\upsilon_{F})$ is shown vs the temperature
parameter $\Delta\left(  T\right)  /T$ for different sets of Landau
parameters. Case (a): $F_{0}=0.2,0.6,1;F_{1}=0.5$. Case (b): $F_{0}%
=0.2,0.6,1;F_{1}=1$. We present the curves for the wave propagation along the
symmetry axis ($\theta_{\mathbf{q}}=0$) and in the perpendicular direction
($\theta_{\mathbf{q}}=\pi/2$).}%
\label{fig2}%
\end{figure}

The density-dependent Landau parameters entering the dispersion equation are
not reliably known. Therefore in Fig. \ref{fig2} we present solutions to Eq.
(\ref{Qsh}) for several sets of the Landau parameters. The curves show the
zero-sound velocity as a function of temperature parameter $\Delta\left(
T\right)  /T$ for a pure $^{3}P_{2}$ pairing (solid curves) and for the case
of pairing into the mixed $^{3}P_{2}-\ ^{3}F_{2}$ state (dashed curves). The
plots are made for the sound propagating along the axis of the wave function
of the condensate ($\theta_{\mathbf{q}}=0$) and in the orthogonal direction
($\theta_{\mathbf{q}}=\pi/2$).

For the case of mixed pairing we have chosen $\Delta_{3}=0.17\Delta_{1}$, in
agreement with that found in realistic calculations by different authors (see,
\textit{e.g.}, Ref. \cite{Khod}). As one can see, a small admixture of the
$^{3}F_{2}$ state does not modify markedly the dispersion curves obtained for
the pure $p$-wave superfluid.

The sound waves are anisotropic. At fixed temperature the velocity of the
sound grows along with deviation of the wave vector from the axis of the order
parameter. The sound speed is maximal for orthogonal propagation.

As regards the temperature dependence, immediately below the critical
temperature $T_{c}$ the velocity of zero sound goes down, and the undamped
wave disappears at some temperature $T_{1}\left(  \theta_{\mathbf{q}}\right)
<T_{c}$ when the sound velocity becomes smaller than the Fermi velocity,
$s_{0}<1$ (although the mode with exponentially small damping can exist in
some region below this temperature). Thus the undamped collective excitation
may or may not exist at some temperature, depending on the values of $F_{0}$
and $F_{1}$. If it does, its velocity $s_{0}\left(  \theta_{\mathbf{q}%
},T\right)  $ will always be greater than unity.

\section{Summary and conclusion}

Let us summarize our results. We have studied the linear response of a
superfluid neutron liquid to an external vector field in the limit
$\omega,q\upsilon_{F}\ll\Delta$. The calculation is made for the case of
$^{3}P_{2}-\,^{3}F_{2}$ condensate which is expected to exist in the
superdense core of neutron stars due to spin-orbit and tensor pairing
interactions. By analyzing the poles of the longitudinal response we have
found the low-energy spectrum of sound-like collecive excitations caused by
density fluctuations in the condensate. Previously the sound-like excitations
were investigated for a triplet condensate caused by central pairing forces in
$^{3}He$ \cite{Wolfe73, Wolfe, Maki}. Our dispersion equation (\ref{Qsh})
represents a generalization of the above results to the case of pairing caused
by noncentral spin-orbit and tensor interactions and naturally recovers
previous results obtained for the case of central forces.

The sound-like spectrum of a Fermi liquid substantially depends on the
residual particle-hole interactions which are conventionally described by a
set of Landau parameters. We have limited our consideration to the first two
terms of this expansion. This approach can be considered as a model of the
Fermi-liquid interactions, although there are indications that the
higher-order Landau interactions do not affect the longitudinal response
functions \cite{Pines, Larkin}.

Unfortunately, the density-dependent Landau parameters entering the dispersion
equation are not reliably known for an asymmetric nucleon matter (although, in
principle, these can be evaluated theoretically \cite{Lehr, Fromel, Dalen,
Lenske}). Therefore we have studied solutions to Eq. (\ref{Qsh}) for several
sets of the Landau parameters. We found that the sound waves are anisotropic
and a small admixture of the $^{3}F_{2}$ state does not modify markedly the
dispersion curves obtained for the pure $p$-wave superfluid neutrons. At fixed
temperature the velocity of the sound grows along with deviation of the wave
vector from the axis of the order parameter. The sound speed is maximal for
orthogonal propagation.

Immediately below the critical temperature $T_{c}$ the velocity of zero sound
decreases when the temperature goes down, and the undamped wave disappears at
some temperature $T_{1}\left(  \theta_{\mathbf{q}}\right)  <T_{c}$ when the
sound velocity becomes smaller than Fermi velocity. (although the mode with
exponentially small damping can exist in some region below this temperature).
Thus the undamped collective excitation may or may not exist at some
temperature, depending on the values of $F_{0}$ and $F_{1}$. If it does, its
velocity will always be greater than Fermi velocity.

In our analysis, we have assumed that the axis of the order parameter is
equally oriented everywhere. It is necessary to notice, however, that texture
effects can orient different parts of the sample differently, and therefore
give a range of frequency shifts which together appear as a broad line. The
texture effects can be minimized by an external magnetic field, although the
magnetic field has no consequence on the dispersion of the sound wave except
that it serves to fix the relative orientation of the spin-orbital wave
function associated with the order parameter, if the dipole interaction is
taken into account.

On the other hand the texture effects could play an important role in neutrino
cooling of neutron star at the latest stage. Indeed, the sound wave can emit a
neutrino pair through neutral weak currents while crossing the border where
the axis changes its direction. Neutrino radiation is possible also due to
collisions of sound waves \cite{Bed}. As already mentioned in the
Introduction, the sound waves are known to play an important role also in a
superfluid heat conduction when the transverse electron motion is strongly
suppressed by a magnetic field \cite{Pons}. Various applications of the
results obtained in this paper will be considered elsewhere.

All the results of this paper depend on the Fermi-liquid functions $F_{0}$,
$F_{1}$ which parametrize the normal Fermi liquid. The only way to evaluate
these functions for a superdense asymmetric nuclear matter is to estimate
their values from first-principles calculations. Although, in practice, such
work is in progress \cite{Lehr, Fromel, Dalen, Lenske}, the complete
information on the Landau parameters for a neutron matter is not available at
the moment. Only the well-known conditions \cite{Migdal} of the matter
stability with respect to long-wave perturbations can be used in order to
limit the Landau parameters.


\begin{thebibliography}{99}                                                                                               %
\bibitem {Tamagaki}R. Tamagaki, Prog. Theor. Phys. 44, 905 (1970).

\bibitem {Hoffberg}M. Hoffberg, A. E. Glassgold, R. W. Richardson and M.
Ruderman, Phys. Rev. Letters 24, 775 (1970).

\bibitem {Takatsuka}T. Takatsuka, Prog. Theor. Phys. 48, 1517 (1972).

\bibitem {Baldo}M. Baldo, J. Cugnon, A. Lejeune and U. Lombardo, Nucl. Phys. A
536, 349 (1992).

\bibitem {Elg}\O . Elgar{\o }y, L. Engvik, M. Hjorth-Jensen, E. Osnes, Nucl.
Phys. A 607, 425 (1996).

\bibitem {Khod}V. V. Khodel, V. A. Khodel, and J. W. Clark, Nucl. Phys. A 679,
827 (2001).

\bibitem {Dean}D. J. Dean and M. Hjorth-Hensen, Rev. Mod. Phys. 75, 607 (2003).

\bibitem {Khodel}M.V. Zverev, J. W. Clark, and V. A. Khodel, Nucl. Phys. A
720, 20 (2003).

\bibitem {L10}L. B. Leinson, Phys. Rev. C 81, 025501 (2010).

\bibitem {L10a}L. B. Leinson, Phys. Lett. B 689, 60 (2010).

\bibitem {L10b}L. B. Leinson, Phys. Rev. C 82, 065503 (2010).

\bibitem {Pons}D. N. Aguilera, V. Cirigliano, J. A. Pons, S. Reddy, and R.
Sharma, Phys. Rev. Lett. 102, 091101 (2009).

\bibitem {Bed}P. F. Bedaque, G. Rupak, and M. J. Savage, Phys. Rev. C 68,
065802 (2003).

\bibitem {Nambu}Y. Nambu, Phys. Rev. 117, 648 (1960).

\bibitem {LP06}L. B.Leinson and A. P\'{e}rez, Phys. Lett. B 638, 114 (2006).

\bibitem {L01}L. B. Leinson, Nucl. Phys. A 687, 489 (2001).

\bibitem {Migdal}A. B. Migdal, \textit{Theory of Finite Fermi Systems and
Applications to Atomic Nuclei} (Interscience, London, 1967).

\bibitem {Gusakov}M. E. Gusakov, Phys. Rev. C 81, 025804 (2010).

\bibitem {Leggett}A. J. Leggett, Phys. Rev. 147, 119 (1966).

\bibitem {Wolfe73}P. W\"{o}lfe, Phys. Rev. Lett. 30, 1169 (1973).

\bibitem {Wolfe}P. W\"{o}lfe, Phys. Rev. Lett. 31, 1437 (1973).

\bibitem {Maki}H. Ebisawa and K. Maki, Prog. Theor. Phys. 51, 337 (1974).

\bibitem {L08}L. B. Leinson, Phys. Rev. C 78, 015502 (2008).

\bibitem {Leggett75}A. J. Leggett, Rev. Mod. Phys. 47, 331 (1975).

\bibitem {Leggett65}A. J. Leggett, Phys. Rev. 140, A1869 (1965).

\bibitem {Pines}D. Pines and P. Nozi\`{e}res, \textit{Theory of Quantum
Liquids} (Benjamin, New York, 1966).

\bibitem {Larkin}A. I. Larkin and A. B. Migdal, Zh. Experim. i Teor. Fiz. 44,
1703 (1963) [Sov. Phys. JETP 17, 1146 (1963)].

\bibitem {Schr}J. Schrieffer, \textit{Theory of Superconductivity} (W.
Benjamin, New York, 1964), p. 157.

\bibitem {Lehr}J. Lehr, M. Effenberger, H. Lenske, S. Leupold and U. Mosel,
Phys. Lett. B483, 324 (2000).

\bibitem {Fromel}F. Fr\"{o}mel, H. Lenske, U. Mosel, Nucl. Phys. A723, 544 (2003).

\bibitem {Dalen}E. N. E. van Dalen, C. Fuchs, and A. Faessler, Phys. Rev.
Lett. 95, 022302 (2005).

\bibitem {Lenske}P. Konrad, H. Lenske, U. Mosel, Nucl.Phys.A 756, 192 (2005).
\end{thebibliography}
\end{document}